\documentclass[preprint,nofootinbib,11pt]{revtex4}
\usepackage{amssymb,amsopn}
\def\h{\hat}
\def\hp{\hat{\partial}}
\def\x{\hat x}
\def\pat{\partial}

\newcommand\ee{\end{equation}}
\newcommand{\ba}[1]{\begin{eqnarray}\label{#1}}
\newcommand{\baa}{\begin{eqnarray}}
\newcommand\ea{\end{eqnarray}}
\newcommand{\be}{\begin{equation}}
\begin{document}
\title{Field theory on kappa-spacetime}  % use lower case
%
          % Leave the remaining items untouched.
\author{Marija Dimitrijevi\'c$^{a,b}$, Larisa Jonke$^{c}$, Lutz M\"oller$^{b,d}$, 
Efrossini Tsouchnika$^d$, Julius Wess$^{b,d}$, Michael Wohlgenannt$^e$}      

\vspace{1cm}

\affiliation{a) University of Belgrade, Faculty of Physics,
Studentski trg 12, 11000 Beograd, Serbia and Montenegro \\
b) Max-Planck-Institut f\"ur 
Physik,
F\"ohringer Ring 6, 80805 M\"unchen, Germany \\ c) Rudjer Boskovic Institute,
Theoretical Physics Division,
PO Box 180, 10002 Zagreb, Croatia \\ d) Universit\"at M\"unchen,
Fakult\"at f\"ur Physik,
Theresienstr.\ 37, 80333 M\"unchen, Germany\\ e) Technische Universit\"at Wien, 
Institut f\"ur Theoretische Physik, Wiedner Hauptstr. 8-10, 1040 Wien, Austria}
%
%
%\pacs{11.10.Nx}     % max. 2 codes of Physics and Astronomy 
               % Classification Scheme
%\keywords{field theory, quantum group, deformation} % lowercase letters
%\maketitle

\vspace{1cm}

\begin{abstract}
A general formalism is developed that allows the construction of 
field theory on quantum spaces which are deformations of ordinary
spacetime. The symmetry group of spacetime
is replaced by a quantum group. This
formalism is demonstrated for the $\kappa$-deformed Poincar\'e algebra and
its quantum space. The algebraic setting is mapped to the algebra of
functions of commuting variables with a suitable
$\star$-product. Fields are elements of this function algebra. 
As an example, the Klein-Gordon equation is defined and   
derived from an action. 
\end{abstract}
\maketitle

\section{Introduction}

Although quantum field theory is extremely successful, 
the combination of general relativity and quantum mechanics suggests 
that spacetime might not be a differential manifold. 
Relying on the well-developed mathematical concept of deformation, 
we formulate a field theory defined on a quantum space rather 
than on the usual differential manifold~\cite{prvi,drugi, treci}. 
The main idea is the following: A differential manifold 
can be described by the algebra of functions on the manifold. 
We deform the usual algebra of functions on Minkowski spacetime 
to obtain the functions on the $\kappa-$Minkowski spacetime. 
In constructing the theory, we implement the $\kappa-$Poincare 
algebra as a deformed symmetry. Physical fields are those 
functions which are representations of the deformed symmetry algebra. 
After defining all field-theory ingredients in an abstract algebra formalism, 
we use the $\star$-product representation to establish a 
connection with usual field theory.
We regard the effective ("noncommutative") action obtained in this way 
as  a  smooth deformation of the standard theory, where a small
parameter of 
deformation should be determined by experimental input.
For a different interpretation see Ref.\cite{amelino} and 
the contributions by N. R. Bruno,  and
by F. J. Herranz in this Proceedings.

\section{Algebraic setting}

The $\kappa-$deformed space is the factor space
of the algebra freely generated by 
$n$ coordinates $\hat{x}^1\dots \hat{x}^n$,
divided by the ideal generated by commutation relations:
\be
\label{1.3}
[\h{x}^k,\h{x}^l] = 0, \;
[\h{x}^n,\h{x}^l]=ia\h{x}^l,\quad  k,l=1,\dots n-1.
\ee
We work in the Euclidean space, where we  rotate the deformation 
vector $a^{\mu}$ into the $n$-th direction\footnote{
The deformation parameter $a=a^n$
is related
to the more common $\kappa$ through $\sqrt{a^2}=\kappa^{-1}$.}.

Derivatives on an algebra have been introduced in \cite{calculus}. They
generate a map in the coordinate space,
elements of the coordinate space are mapped to other
elements of the coordinate space. Thus, they have to be consistent
with the algebra relations and for $a =0$, they should behave
like ordinary derivatives.
In addition, they should act
at most linearly in the
coordinates and  the derivatives, and commute among themselves.
These requirements are
satisfied by the following rules for differentiation\footnote{This
solution is not unique, but the ambiguity does not show up in
the physical action, see Ref.\cite{treci}}:
\be
\label{1.8}
\left[\hp _n, \hat x^\mu\right] = \delta_n^\mu,\;
\left[\hp _i, \hat x^\mu\right] = \delta_j^\mu+ia\delta^\mu_n\hp_i.
\ee
Derivatives in the $k$-th direction have the deformed  Leibniz rule:
\be
\h{\partial}_k(\hat{f}\cdot \hat{g})= (\h{\partial}_k\hat{f})\cdot \hat{g}
+(e^{ia\hp_n}\h{f})\cdot \h{\partial}_k\hat{g} .\label{1.8a}
\ee

The symmetry structure of the space is a deformation of the $n$-dimensional 
group of rotations. The generators of symmetry are maps of the 
coordinate space 
consistent with the relations (\ref{1.3}):
\begin{eqnarray}
\label{1.4}
\left[ M^{kl},\hat x^\mu \right] &=& \delta^{k\mu} \hat x^l 
- \delta^{l\mu} \hat x^k ,
\nonumber \\
\left[ M^{kn}, \hat x^\mu \right] &=& \delta^{k\mu} \hat x^n
- \delta^{n\mu} \hat x^k -ia M^{k\mu}. 
\end{eqnarray}
From (\ref{1.4}) it is possible to compute the commutators
of the
generators: 
\be
\label{1.7}
\left[ M^{\mu\nu},M^{\rho\sigma}\right] = 
\delta^{\mu\rho}M^{\nu\sigma}+\delta^{\nu\sigma} M^{\mu\rho}
-\delta^{\mu\sigma} M^{\nu\rho} 
- \delta^{\nu\rho}M^{\mu\sigma}.
\ee
This is the undeformed $SO(n)$ algebra,
but the comultiplication is deformed for  generators of 
rotations involving the $n$-th direction:
\be
\label{comult}
M^{kn}(\h f\cdot\h g)=\left(M^{kn}\h f\right)\cdot\h g+
\left(e^{ia\hp_n}\h f\right)\cdot M^{kn}\h g+i a 
\left(\hp_l\h f\right)\cdot M^{kl}\h g.
\ee

The derivatives introduced in (\ref{1.8}) 
have complicated transformation properties 
under rotation, see Ref.\cite{prvi}. For physical applications
we construct
the  derivatives 
with the usual, undeformed transformation properties under rotation.
A deformed Laplace operator (see Refs.\cite{lukrue}) and a deformed
Dirac operator (see Refs.\cite{Dirac}) can be defined. For
the Laplace operator $\hat \Box$,  we demand that it should commute
with the generators of the algebra
$\left[ M^{\mu\nu},\hat{\Box}\right] =0$, 
and that it should be a deformation of the usual Laplace
operator. By iteration in $a$ we find
\begin{equation}
\hat \Box = e^{-ia\hp _n} \hat{\Delta} +{2 \over a^2}
\Big( 1 - \cos ( a\hp _n ) \Big). \label{1.12}
\end{equation}

Since the $\gamma$-matrices are $\hat{x}$-independent and
transform as usual, the covariance of the full Dirac operator
$\gamma^\mu \hat{D}_\mu$ implies
that the transformation law of its components is vector-like:
\be\label{dir}
\left[ M^{\mu\nu},\hat{D}_\rho\right] = \delta _\rho^\mu\hat{D}^\nu
-\delta _\rho^\nu\hat{D}^\mu.\ee
These relations are obviously consistent with the algebra
(\ref{1.7}).
The components of Dirac  operator that satisfy
(\ref{dir}) and  have the correct limit for $a \rightarrow 0$ are
\begin{eqnarray}
\hat{D}_n &=& {1\over a} \sin (a\hp _n)+
{ia\over 2}\hat{\Delta}\hspace{1mm} e^{-ia\hp _n}, \nonumber \\
\hat{D}_i &=&\hp _i e^{-ia\hp _n}. \label{1.14}
\end{eqnarray}

Physical fields are formal power-series expansions
in the coordinates and as such are elements of the
coordinate algebra:
\begin{equation}
\label{3.1}
\hat \phi(\hat x) = \sum _{ \{\alpha\} } c_{\alpha_1\dots\alpha_n} \, :
(\x^1)^{\alpha_1}\dots (\x^n)^{\alpha_n}:.
\end{equation}
The summation is over a basis in the
coordinate algebra, as indicated by colons.
The field can also be defined by its coefficient functions
$c_{\{\alpha_1\dots\alpha_n\}}$, once the basis is specified.
Fields can be added, multiplied, differentiated
and transformed. 
For example, we define the transformation law of a scalar field as
$$\hat \phi(\hat x) =(1+\varepsilon_{\mu\nu}M^{\mu\nu})\hat{\phi}'(\hat x).$$
Because of the nontrivial coproduct of the $M^{kn}$ generator (\ref{comult}), 
we 
cannot use the usual $\phi'(x')=\phi(x)$ definition.

Having defined all  ingredients, we can write
the equation of motion for a
free scalar field, invariant under the action of the symmetry generators by 
construction:
\be
\label{eom}
\left( \hat \Box + m^2 \right) \hat\phi(\x) = 0.
\ee

\section{The $\star$-product}

The framework of deformation quantization \cite{def}, allows to map the 
associative algebra of functions on noncommutative space to an algebra 
of functions on a commutative space by means of $\star$-product.
In short, the idea is as follows:
We consider polynomials of fixed degree in the algebra
- homogeneous polynomials. They form a finite-dimensional
vector space. For an  algebra with  the Poincar\' e-Birkhoff-Witt
property (and a Lie algebra has this property),
the dimension of the vector space of homogeneous polynomials
in the algebra is the same as for polynomials of
commuting variables. Thus, there is an isomorphism
between two finite-dimensional  vector spaces. This
vector space isomorphism can be extended to an algebra
isomorphism by defining the product of polynomials of
commuting variables by first mapping these polynomials back
to the algebra, multiplying them there and mapping the product
to the space of polynomials of ordinary variables. The product we obtain
in this way is called $\star$-product. It is noncommutative and
contains the information about the product in the algebra.

An efficient way of computing the $\star$-product is  Weyl quantization.
Although one can find a closed form 
for the $\star$-product (see Ref.\cite{treci}), it is more
instructive to write an expanded expression, up to second order in 
deformation parameter $a$:
\begin{eqnarray}
\label{2.4}
f \star g\, (x) & = & f(x)g(x) + {ia\over 2} x^j \Big( \pat_n f(x) \pat_j g(x)
        -  \pat_j f(x) \pat_n g(x) \Big) \nonumber\\
	& & -{a^2\over 12}x^j \Big( \pat_n^2 f(x) \pat_j g(x)-\pat_j\pat_n f(x) \pat_n g(x)\nonumber\\
	&  &\qquad\qquad -\pat_n f(x) \pat_j \pat_n g(x)+ \pat_j f(x) \pat_n^2 g(x)\Big)  \nonumber\\
	&  & -{a^2\over 8} x^j  x^k\Big(\pat_n^2 f(x) \pat_j\pat_k g(x) - 2
	        \pat_j \pat_n f(x) \pat_n\pat_k g(x)\\
		&  &\qquad\qquad
		        +\pat_j \pat_kf(x) \pat_n^2 g(x)
			        \Big) + \mathcal{O}(a^3). \nonumber
				\end{eqnarray}

From the action of  an operator $\hat O$ on 
symmetric polynomials  in the algebra we  compute the
action of an operator $O^*$ on  ordinary functions.
For example,
\begin{eqnarray}
  \label{2.13}
  \pat_i^* f(x) & = & \pat_i {e^{ia\pat_n} -1\over ia\pat_n} \, f(x),\nonumber \\
 M^{*ln}   f(x) & = & \left( x^l\pat_n - x^n\pat_l+ x^l\pat_\mu\pat_\mu 
  \frac{ e^{ia \pat_n}-1}{2\pat_n } - x^\nu\pat_\nu\pat_l 
  \frac{e^{ia\pat_n} -1- ia\pat_n}{ia\pat_n^2} \right) f(x)\nonumber.
  \end{eqnarray}
The operators  inherit the Leibniz rule from the algebra.
 
Now we can write the Klein-Gordon  equation of motion (\ref{eom}) 
in the following form:
\begin{equation}
\label{3.11}
\left( \Box^* + m^2 \right) \phi(x) = \left(
-{2\over a^2 \pat_n^2} (\cos(a\partial_n) - 1)\Box
 + m^2 \right) \phi(x)=0.
\end{equation}
Expanding the equation in $a$ will give us the equation of motion for 
the free scalar field with second-order correction:
\be
\left(\Box+m^2-\frac{a^2\pat_n^2}{12}\Box+ \mathcal{O}(a^3)\right)\phi(x)=0.\ee

\section{The Variational Principle}

We derive field equations by means of a variational principle
such that the dynamics can be formulated with the help of
the Lagrangian formalism. For this purpose, we need an
integral. 
We  define it in the $\star$-product formalism
and use the usual definition of an integral
of functions of commuting variables. 
Such an integral
in
general  will not have the trace property, but we can introduce
 a measure function
to achieve it:
\begin{equation}
\label{6.4}
\int\textrm{d}^n x \hspace{1mm}\mu(x)\hspace{1mm}(f(x)\star g(x)) =\int\textrm{d}^n x\hspace{1mm} \mu(x)\hspace{1mm}(g(x)\star f(x)).
\end{equation}
Note that $\mu(x)$ is not $\star$-multiplied with the other
functions, it is part of the volume element.
Using Eq.(\ref{6.4}) as a definition of the measure function, 
and Eq.(\ref{2.4}),
 we  obtain
$$\partial_n \mu(x) =0, \qquad x^j \partial_j \mu (x)= (1-n)\mu(x).$$
We also have to define "improved" differential operators $\mathcal{O}$
which are hermitean in the sense
\begin{equation}
\label{6.15}
\int\textrm{d}^n x\hspace{1mm} \mu\hspace{1mm}\bar{f}\star
\mathcal{O}g =  \int\textrm{d}^n x\hspace{1mm}
\mu\hspace{1mm}\overline{\mathcal{O}f}\star g.
\end{equation}
This is achieved by the following redefinition of all differential operators:
\begin{equation}
\label{6.18}
\partial^*_k\rightarrow {\tilde \partial}^*_k = \left(\partial_k
+\frac{\partial_k\mu}{2\mu}\right)\frac{e^{ia\partial_n}-1}{ia\partial_n}.
\end{equation}

Now we define the variational principle in such a way that the function to be 
varied is brought to the left by cyclic permutation and varied:
\begin{equation}
\label{6.10a}
\frac{\delta}{\delta g}\int\textrm{d}^n x\;
\mu\hspace{1mm} f\star g \star h = 
\frac{\delta}{\delta g}\int\textrm{d}^n x\hspace{1mm}
\mu\hspace{1mm}  g  (h \star f) = \mu\hspace{1mm} (h\star f).
\end{equation}

With this definition, after performing a suitable field 
redefinition~\cite{prvi},
we  derive the equation of motion (\ref{3.11}) from the 
following action: 
\be\label{act}
S=\frac{1}{2}\int \textrm{d}^n x\mu\hspace{1mm}\phi(x)
\star\left( \tilde{\Box}^* + m^2 \right) \phi(x).\ee
The operator $\tilde{\Box}^*$ is the improved
Laplace operator $\Box^*$ in the sense of (\ref{6.18}).

\section{Outlook}

Using the formalism developed in Ref.\cite{prvi} and presented here, one
can also construct  gauge theories on $\kappa$-spacetime, see Ref.\cite{drugi}.
The main consequences of deformation of the coordinate algebra for
gauge theories are:\\
a) Gauge fields are enveloping-algebra-valued, and, therefore, one must 
construct a (Seiberg-Witten) map to restrict the theory to the finite 
(Lie-algebra) 
number of degrees of freedom.\\
b) Gauge fields are derivative-valued, as  a consequence of deformed 
Leibniz rules.

An important open problem is the construction of an invariant action. 
Namely, we can construct an 
action invariant under gauge transformation OR an action 
invariant under the action of 
symmetry generators (using the "quantum trace" instead of the 
integral defined in 
(\ref{6.4}), see Ref.\cite{lt}).
The work on the  problems of quantization and on the formulation of 
"deformed" conservation laws is in progress. 

\section*{Acknowledgments}
L.J. would like to thank the organizers for the  very interesting and
pleasant workshop.
This work is supported by the Ministry of Science and Technology
of the Republic of Croatia under the contract 0098003, and partially by 
the Alexander
von Humboldt Foundation.

\end {document}